%% file: main.tex
\documentclass[aip,jcp,reprint]{revtex4-1}

\usepackage[utf8]{inputenc} 
\usepackage[T1]{fontenc}    
\usepackage{hyperref}       
\usepackage{url}            
\usepackage{booktabs}       
\usepackage{amsfonts}       
\usepackage{amsmath}
\usepackage{nicefrac}       
\usepackage{microtype}      
\usepackage{graphicx}
\usepackage{lipsum}
\usepackage{comment}
\usepackage[caption=false]{subfig}
\usepackage{textcomp}
\usepackage{soul}           
\usepackage{amsmath}
\usepackage{listings}
\usepackage{chngcntr}
\usepackage{enumitem}
\usepackage[table,xcdraw]{xcolor}
\usepackage{booktabs}
\usepackage{multirow}
\usepackage{hhline}
\usepackage{amssymb}
\usepackage{siunitx}
\usepackage[version=3]{mhchem}
\usepackage{euler}
\usepackage{bm}

\definecolor{codegreen}{rgb}{0,0.6,0}
\definecolor{codegray}{rgb}{0.5,0.5,0.5}
\definecolor{codepurple}{rgb}{0.58,0,0.82}
\definecolor{backcolour}{rgb}{0.95,0.95,0.92}

\lstdefinestyle{mystyle}{
    backgroundcolor=\color{backcolour},   
    commentstyle=\color{codegreen},
    keywordstyle=\color{magenta},
    numberstyle=\tiny\color{codegray},
    stringstyle=\color{codepurple},
    basicstyle=\ttfamily\footnotesize,
    breakatwhitespace=false,         
    breaklines=true,                 
    captionpos=b,                    
    keepspaces=true,                 
    numbers=left,                    
    numbersep=5pt,                  
    showspaces=false,                
    showstringspaces=false,
    showtabs=false,                  
    tabsize=2
}
\let\oldtextbf=\textbf
\renewcommand*{\textbf}[1]{\ifmmode\mathbf{#1}\else\oldtextbf{#1}\fi}
\renewcommand*{\phi}[0]{\varphi}

\newcommand{\paragraphtitle}[1]{\textsf{\textbf{\small {#1}}}}

\draft 

\begin{document}

\title{Committors without Descriptors } 



\author{Peilin Kang}
\altaffiliation{These authors contributed equally: Peilin Kang and Jintu Zhang}
\affiliation{Atomistic Simulations, Italian Institute of Technology, 16156 Genova, Italy}

\author{Jintu Zhang}
\altaffiliation{These authors contributed equally: Peilin Kang and Jintu Zhang}
\affiliation{Atomistic Simulations, Italian Institute of Technology, 16156 Genova, Italy}
\affiliation{Innovation Institute for Artificial Intelligence in Medicine of Zhejiang University, College of Pharmaceutical Sciences, Zhejiang University, Hangzhou 310058 Zhejiang, China}

\author{Enrico Trizio}
\affiliation{Atomistic Simulations, Italian Institute of Technology, 16156 Genova, Italy}

\author{TingJun Hou$^*$}
\email[]{tingjunhou@zju.edu.cn}
\affiliation{Innovation Institute for Artificial Intelligence in Medicine of Zhejiang University, College of Pharmaceutical Sciences, Zhejiang University, Hangzhou 310058 Zhejiang, China}
\affiliation{State Key Lab of CAD\&CG, Zhejiang University, Hangzhou, Zhejiang 310058, China}

\author{Michele Parrinello$^*$}
\email[]{michele.parrinello@iit.it}
\affiliation{Atomistic Simulations, Italian Institute of Technology, 16156 Genova, Italy}


\date{\today}

\begin{abstract}
\input{manuscript/abstract}
\end{abstract}


\maketitle 

\input{manuscript/paper}

\section*{Code and Data Availability} \label{sec:code_avail}
The code for the training of the NN-based committor model alongside didactic tutorials is available through the open-source \texttt{mlcolvar} library~\cite{bonati2023mlcolvar}, which is the preferred way to access the most updated code. Training and simulation data and inputs are available on \hyperlink{https://github.com/alphatestK/GNN-Committor}{https://github.com/alphatestK/GNN-Committor}.


\begin{acknowledgments}
This work received support from the National Natural Science Foundation of China (22220102001).
The authors thank Julian Gale and Blake Armstrong for providing the forcefield for the calcium carbonate system.
\end{acknowledgments}

\section*{Bibliography}
\bibliography{references}

\setcounter{section}{0}
\renewcommand{\thesection}{S\arabic{section}}
\setcounter{equation}{0}
\renewcommand{\theequation}{S\arabic{equation}}
\setcounter{figure}{0}
\renewcommand{\thefigure}{S\arabic{figure}}
\setcounter{table}{0}
\renewcommand{\thetable}{S\arabic{table}}
    
\clearpage
\onecolumngrid

{\Large\normalfont\sffamily\bfseries{{Supporting Information}}}

\setlength{\tabcolsep}{18pt}
\renewcommand{\arraystretch}{1.2}

\input{manuscript/supporting}


\end{document}

%% file: manuscript/abstract.tex
The study of rare events is one of the major challenges in atomistic simulations, and several enhanced sampling methods towards its solution have been proposed.
Recently, it has been suggested that the use of the committor, which provides a precise formal description of rare events, could be of use in this context.
We have recently followed up on this suggestion and proposed a committor-based method that promotes frequent transitions between the metastable states of the system and allows extensive sampling of the process transition state ensemble. 
One of the strengths of our approach is being self-consistent and semi-automatic, exploiting a variational criterion to iteratively optimize a neural-network-based parametrization of the committor, which uses a set of physical descriptors as input.
Here, we further automate this procedure by combining our previous method with the expressive power of graph neural networks, which can directly process atomic coordinates rather than descriptors.
Besides applications on benchmark systems, we highlight the advantages of a graph-based approach in describing the role of solvent molecules in systems, such as ion pair dissociation or ligand binding.

%% file: manuscript/paper.tex
\section{Introduction}

Atomistic simulations are an indispensable tool in the study of complex physicochemical processes. 
However, such simulations find one of their limits in the gap between the affordable simulation time and the typically much longer timescale over which many important phenomena like chemical reactions, protein folding, and crystallization take place. 
Such processes are indeed characterized by rare transitions between metastable states, which are separated by large free energy barriers that act as kinetic bottlenecks hindering sampling. 
This has been called the rare event problem, and, since the introduction of umbrella sampling some fifty years ago, a multitude of enhanced sampling approaches have been suggested to solve it.\cite{Henin2022enhanced}   

Recently, we have introduced a new enhanced sampling method based on the committor function that greatly alleviates the rare event problem\cite{trizio20241} by promoting extensive sampling of both transition and metastable states, and have started using this approach to solve a number of real-life problems.\cite{kang2024computing,trizio20241,deng2025role,das2025machine} 
We recall here that the committor  $q(\boldsymbol{x})$  is a function of the atomic coordinates $\boldsymbol{x}$  which, given two metastable states $A$ and $B$, gives the probability that a trajectory started in $\boldsymbol{x}$ reaches $B$ without having first passed by $A$.~\cite{weinan2010transition}
The committor is arguably the most precise way of describing rare events, since it is a quantity that remains well-defined even if the transition from $A$ to $B$ follows different competing pathways or passes through an intermediate metastable state.  
The committor is also believed to be the optimal one-dimensional reaction coordinate.\cite{ma2005automatic} 

Unfortunately, if one follows the committor formal definition, a rather expensive trial-and-error strategy is needed for its determination. \cite{bolhuis2002transition}
Recently, alternative ways have been proposed to learn the committor function from simulation data using machine learning tools in combination with different learning criteria and enhanced sampling schemes\cite{li2019computing, khoo2019solving, rotskoff2022active, mitchell2024committor, kang2024computing, trizio20241, chen2023committor, chen2023discovering, jung20231, megias2025iterative, breebaart2025understanding}. Such methods have been recently  reviewed   in Ref.\citenum{zhu2025enhancedsamplingagemachine}.
Our contribution to this area \cite{kang2024computing, trizio20241} is based on the Kolmogorov variational principle, which is obeyed by the committor ~\cite{kolmogoroff1931analytischen} which has been solved in a self consistent procedure that eventually leads not only to the calculation  of  $q(\boldsymbol{x})$ but to an extensive and balanced evaluation of the free energy equation. 

However, like in all variational calculations, the quality of the results depends on the expressivity of the trial function used  \cite{kang2024computing}.  
In our initial approach, we have represented $q(\boldsymbol{x})$ as a feed-forward neural network $q(\boldsymbol{x}) = q_\theta(\boldsymbol{d}(\boldsymbol{x}))$ whose input is a set of descriptors $\boldsymbol{d}(\boldsymbol{x})$  chosen to be invariant with respect to the symmetries of the system and whose weights $\theta$ are optimized to minimize the functional associated with Kolmogorov variational problem. 
Despite having proven to be effective in several challenging systems\cite{kang2024computing, trizio20241, deng2025role, das2025machine}, this approach still relies, at least partially, on the user's insight for the choice of appropriate descriptors, which, in complex cases, might not be easy to select.

\begin{figure*}[t!]
    \centering
    \includegraphics[width=0.8\textwidth]{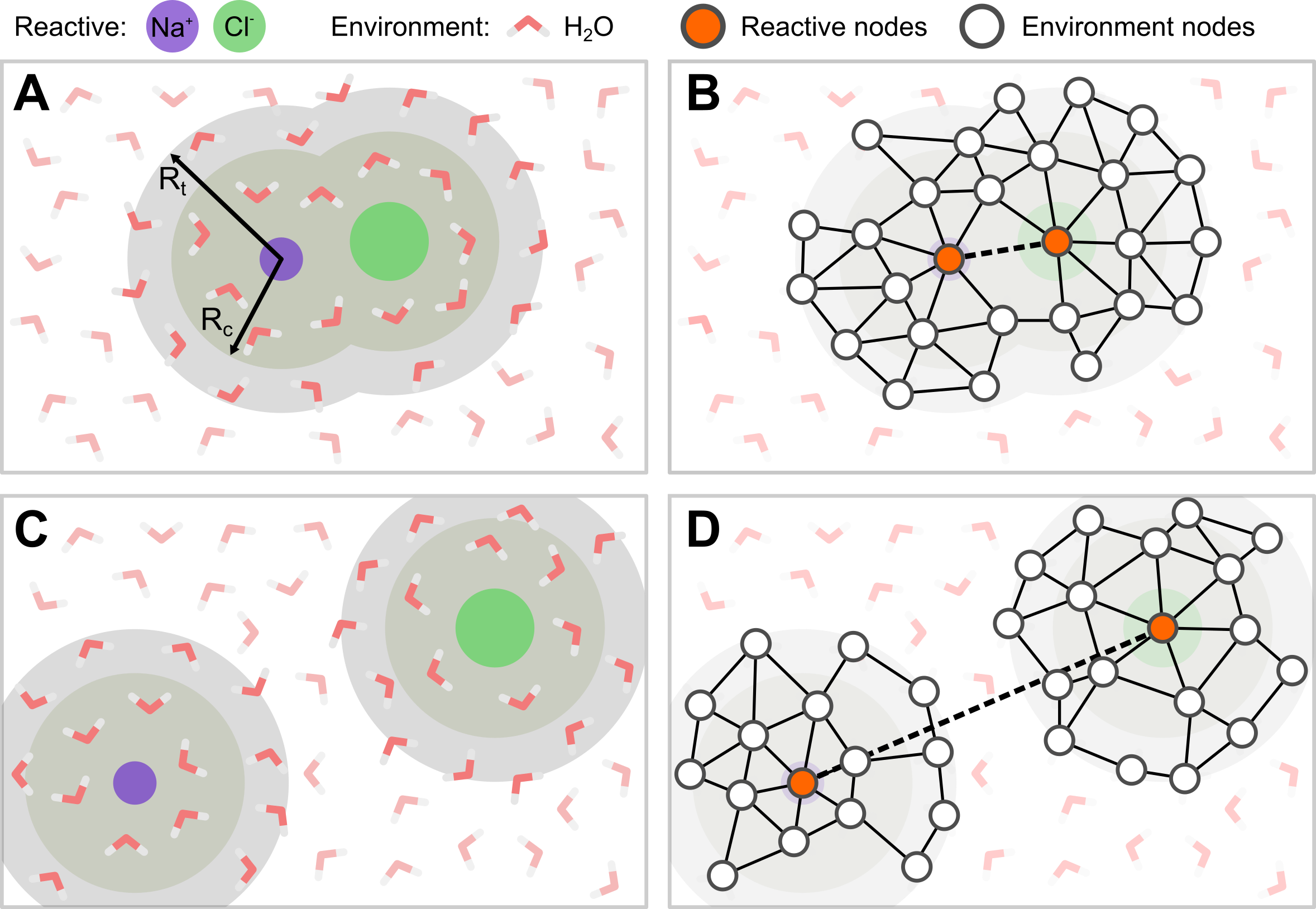}
    \caption{Truncated graph.
    Schematic representation, using the dissociation of \ce{NaCl} as an example, of the truncated graph construction to reduce the computational cost of using GNN-based methods when a few \textit{reacting atoms} (NaCl) interact with a large number of \textit{environment atoms} (H$_2$O).
    Only the atoms belonging to the neighborhood defined by the cutoff radius $R_t$ from reacting atoms are associated with graph nodes, whereas the other are neglected (A and C).
    Such nodes are then connected with edges based on the cutoff $R_c < R_t$ (B and C).
    To avoid having reacting atoms in disconnected graphs, edges between reacting atoms can be enforced regardless of the distance (D).
    The whole truncated graph is processed through the GNN model, but only the reacting atoms are considered in the readout function to obtain the final CV output. }
    \label{fig:truncated_graph}
\end{figure*}

The purpose of this paper is to remove as much as possible this potential obstacle and make the calculation of $q(\boldsymbol{x})$ as automatic as possible.
To achieve this goal, we parametrize the committor with a geometric Graph Neural Network (GNN)~\cite{pietro2023}, which can directly use as input the atomic Cartesian coordinates $\boldsymbol{x}$ while respecting the invariance laws of the system.  
This architecture has already proven to be useful in building machine-learning potentials\cite{batzner2022,grabor2022,cheng2024} and in designing collective variables (CVs)~\cite{speybroeck2024,salvalaglio2024,tiwary20242,zhang2024gnncv,tiwary20243}.
When using a  GNN,  an atomic system is naturally represented by a graph whose nodes are the atoms and whose connecting edges describe their  relationship.  Optionally, nodes and/or edges can be assigned attributes that encode information on the system, greatly facilitating the analysis of the results,\cite{zhang2024gnncv}, especially if an attention mechanism is added to the GNN structure.\cite{li2019attention} 
Furthermore, GNN architectures can be made invariant or equivariant with respect to the symmetry operations of the system.
However, due to their more complex structure, GNN models are  computationally more expensive when compared to standard models, with the overall cost increasing with the complexity and size of the graph.  To alleviate this limitation, we have also introduced new ways of reducing the computational cost of using a GNN when dealing with reactions in condensed phases (see Fig.~\ref{fig:truncated_graph} and Sec.~\ref{sec:methods}). 

In the field of enhanced sampling, it is customary to use as a test the conformational equilibrium of alanine dipeptide. We will stick to this tradition and start with the study of this molecule.
Having passed the dipeptide test, we will demonstrate the usefulness of this new approach in a number of more physically relevant examples in which the solvent plays an important role.
The examples studied are the dissociation of \ce{NaCl} and of \ce{CaCO3} in water, and the binding of an organic molecule to calixarene, a simplified but still 
representative model of drug–protein interaction. All these systems have been studied with other means and thus they provide a good testing ground also for the versatility of the committor since their reactive processes can exhibit multiple reaction pathways and/or metastable intermediate states\cite{mullen2014transmission,armstrong2023solubility,piaggi2024ab,rizzi2021,pampel2022improving}. 
A full description of the methods, a summary of the procedure, and other technical details of GNN architecture are provided in Sec.~\ref{sec:methods} and in the Supporting Information.

\section{Results}
\label{sec:results}

The technical details of the simulations, GNN hyperparameters, and training procedures for the examples presented are provided in the Supporting Information (SI).

   \begin{figure}[h!]
        \centering
        \includegraphics[width=\linewidth]{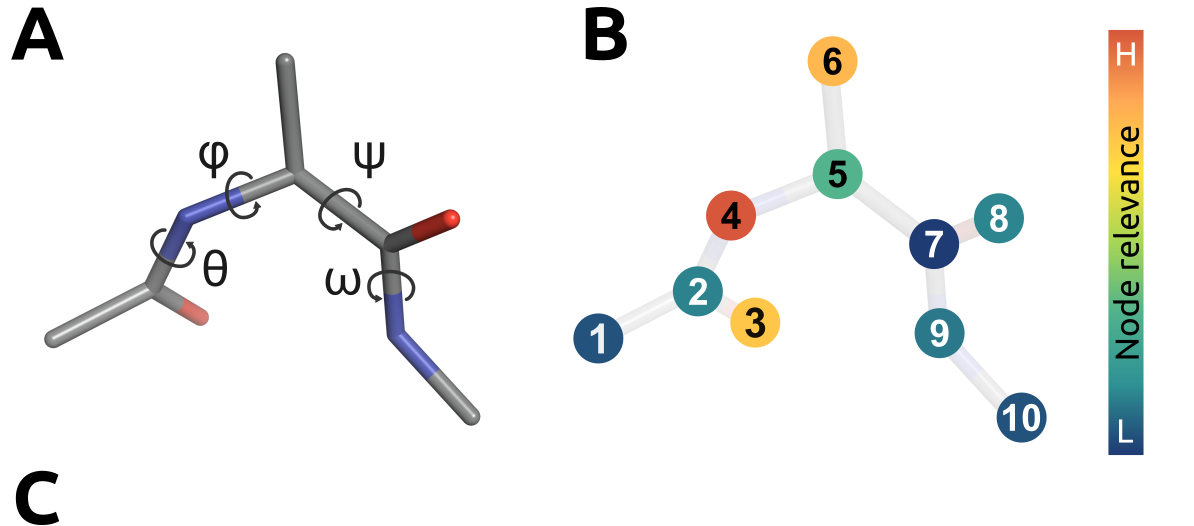}
        \includegraphics[width=\linewidth]{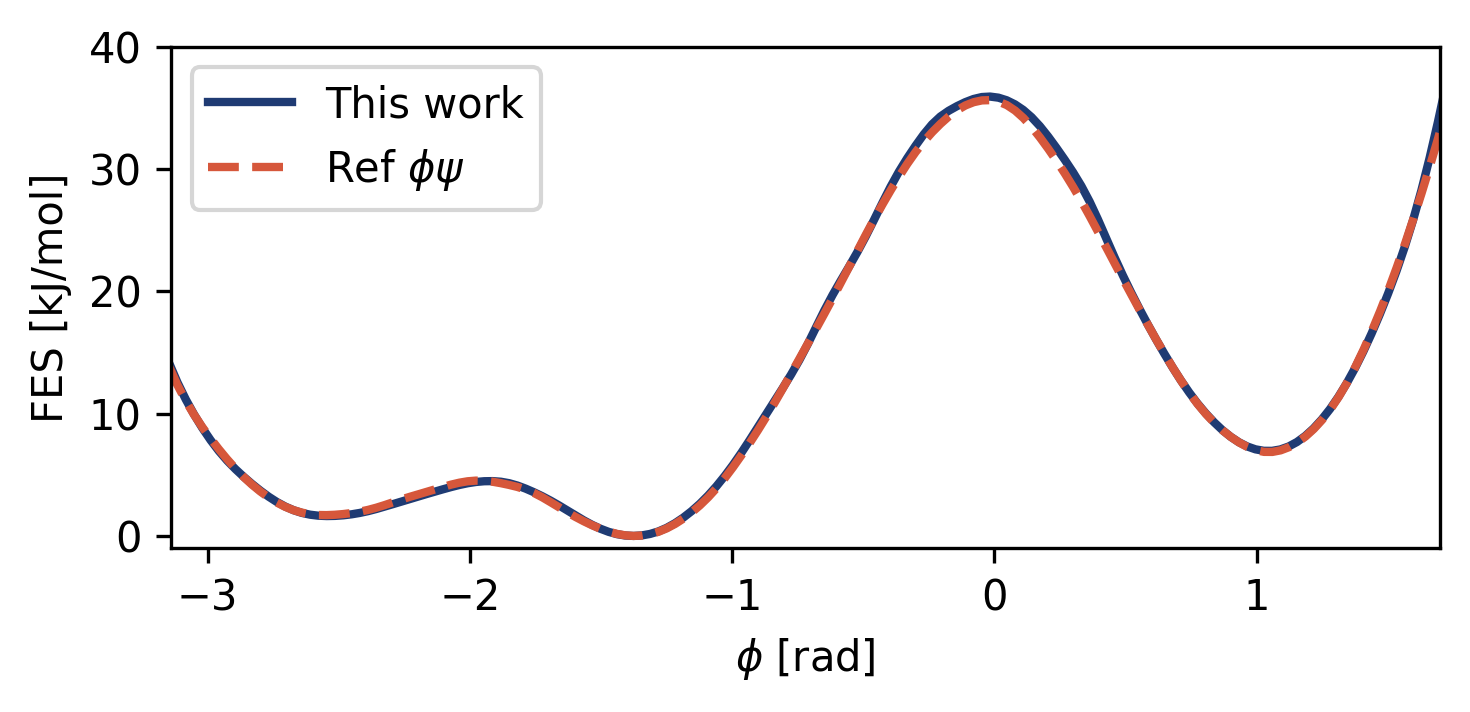}
        \caption{Alanine dipeptide. 
        A) Relevant torsional angles of alanine dipeptide. 
        B) Relative relevance, given by the color scale, to the GNN-based committor model of the graph nodes associated with the heavy atoms.
        C) Free energy surface (FES) projected along the $\phi$ torsional angle obtained with the GNN-based setup proposed in this work (blue solid curve) and a reference OPES simulation using $\phi$ and $\psi$ as CVs (red dashed curve).}
        \label{fig:alanine}
    \end{figure}
    
\subsection{Alanine Dipeptide}

In most studies devoted to rare event sampling, the transition of alanine dipeptide in vacuum between the $C7_{eq}$ ($A$) and $C7_{ax} $ ($B$) conformers is commonly used to demonstrate the effectiveness of new methods. This system is one of the most extensively studied models for rare events and is often characterized by two dihedral angles, \( \phi \) and \( \psi \).
Following Ref.\citenum{kang2024computing}, we begin with unbiased simulation data sampled in the two metastable basins.
To build the input for our GNN model, we consider the 10 heavy atoms of the molecule, which serve as the nodes in the input graph.

Using the optimized GNN model, which reaches convergence within 4 iterations (see SI for details), to drive our enhanced sampling simulations, we obtain an estimate of the free energy surface (FES) in close agreement with reference results obtained by biasing the conventional $\phi, \psi$. 
In addition,  we can automatically identify the key atoms involved in the transition from the GNN architecture. 
In particular, a node sensitivity analysis reveals that atoms 3, 4, 5, and 6 contribute most significantly to the learned committor (see Fig.\ref{fig:alanine}B).
As expected, these are the atoms closely related to the dihedral angles $\phi$ and $\theta$, in full agreement with previous studies.\cite{chandler2000, kang2024computing}

\subsection{Calixarene}

\begin{figure*}[t!]
    \centering
    \begin{minipage}{1.0\linewidth}
        \includegraphics[width=\linewidth]{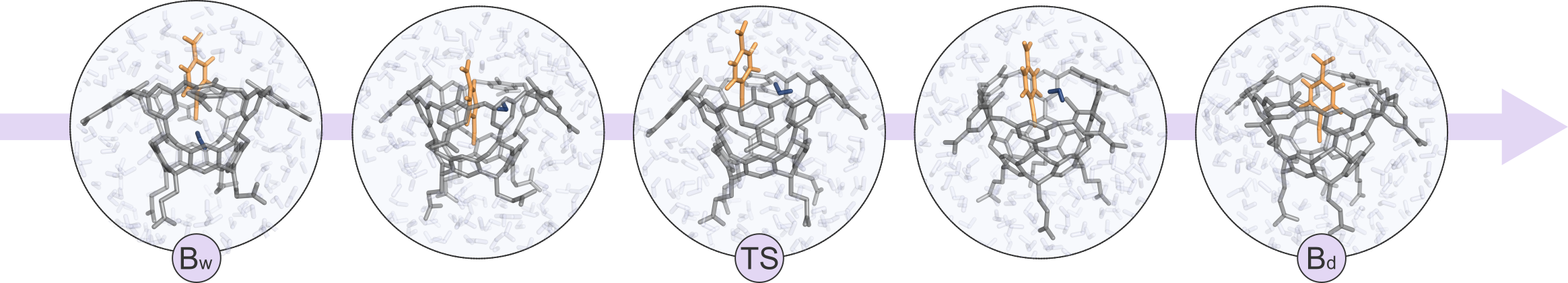} 
    \end{minipage}
    \caption{Calixarene. Snapshots of representative configurations along the dominant reaction pathway of $B_w$ to $B_d$ in the binding process of the \ce{G2} ligand (orange) to the OAMe host molecule (grey) in water, grouped according to the $z$ value.  The water molecules are represented as blue sticks, depicted in transparency when outside the binding cavity and in solid color when inside.}
    \label{fig:Calixarene_snapshots}
\end{figure*}

Having paid homage to tradition, we now study the interaction of the \ce{G2} ligand(4-cyanobenzoic acid) with an octa-acid calixarene host (OAMe) from the \verb|SAMPL5| challenge,~\cite{yin2017overview}, which provides a good example of some features of protein-ligand systems.

Previous studies have shown that the system primarily follows two transition pathways: a "wet"  path, in which the process proceeds through the semi-bound state \( B_{w} \) before reaching the fully-bound state \( B_{d} \), and a ``dry'' path, where guest molecules enter and exit the pocket directly.  
In addition, the wet path has been shown to be dominant even if it involves more intricate dynamics.
Here, we shall not repeat the calculation of Ref.\cite{trizio20241}, but focus only on one of the rate-limiting step, i.e., the transition from the \( B_{w} \) to \( B_{d} \) with the purpose of illuminating the role of water in this transition. 
The results obtained with the setup presented in this work for the whole process are available in the SI, including the overall free energy surface and the estimate for the binding energy.

For the \( B_{w} \) to \( B_{d} \) rate-limiting step, our algorithm finds two pathways, one in which the ligand is solvated while at the same time the calixarene is emptied of water and eventually binds to the now dry calixarene. (see SI Fig. \ref{SIfig:Calixarene_snapshots_twopath})
However, as the contribution of this path to the rate is negligibly small (< 5$\%$, based on analysis of $p_{\mathcal{K}}$ contributions), we consider here only the dominant path (see Fig.\ref{fig:Calixarene_snapshots}). 
In this dissociation mode, the water that is trapped in the calixarene in the \( B_{w} \) state accompanies the ligand to the entrance of the calixarene until it is released into the bulk solvent, leaving room for the ligand to bind to its lowest free energy state.

\subsection{\ce{NaCl}}

We now discuss the case of NaCl dissociation in water, in which the solvent clearly plays a central role. 
The contact ion pair (CIP) is our state $A$, while $B$ is the dissociated state.   
To focus on the dissociation process itself, we set an artificial repulsive wall that limits the inter-ionic distance at 6 \unit{\angstrom}. This implies that in the calculation of the free energy of the $B$ state, part of the entropic contribution is missing.
    
In our previous work\cite{kang2024computing, trizio20241} and also in the case of alanine dipeptide, we started the committor learning self-consistent procedure with two unbiased simulations started in $A$ and $B$, i.e., without any data whatsoever coming from the transition state region.  
Despite being a procedure of general applicability, this is often an overkill since one can often obtain some information on the TS by running first a metadynamics-like simulation, even if driven by a suboptimal CV.  
To showcase this strategy, we start with data coming from an OPES simulation in which the inter-ionic distance was used as CV.  
Even for a simple example like this, this approach allowed reducing the number of iterations needed to reach convergence (see SI, Sec.~\ref{sup_sec:Nacl training}).

\begin{figure}[b!]
        \centering
            \includegraphics[width=\linewidth]{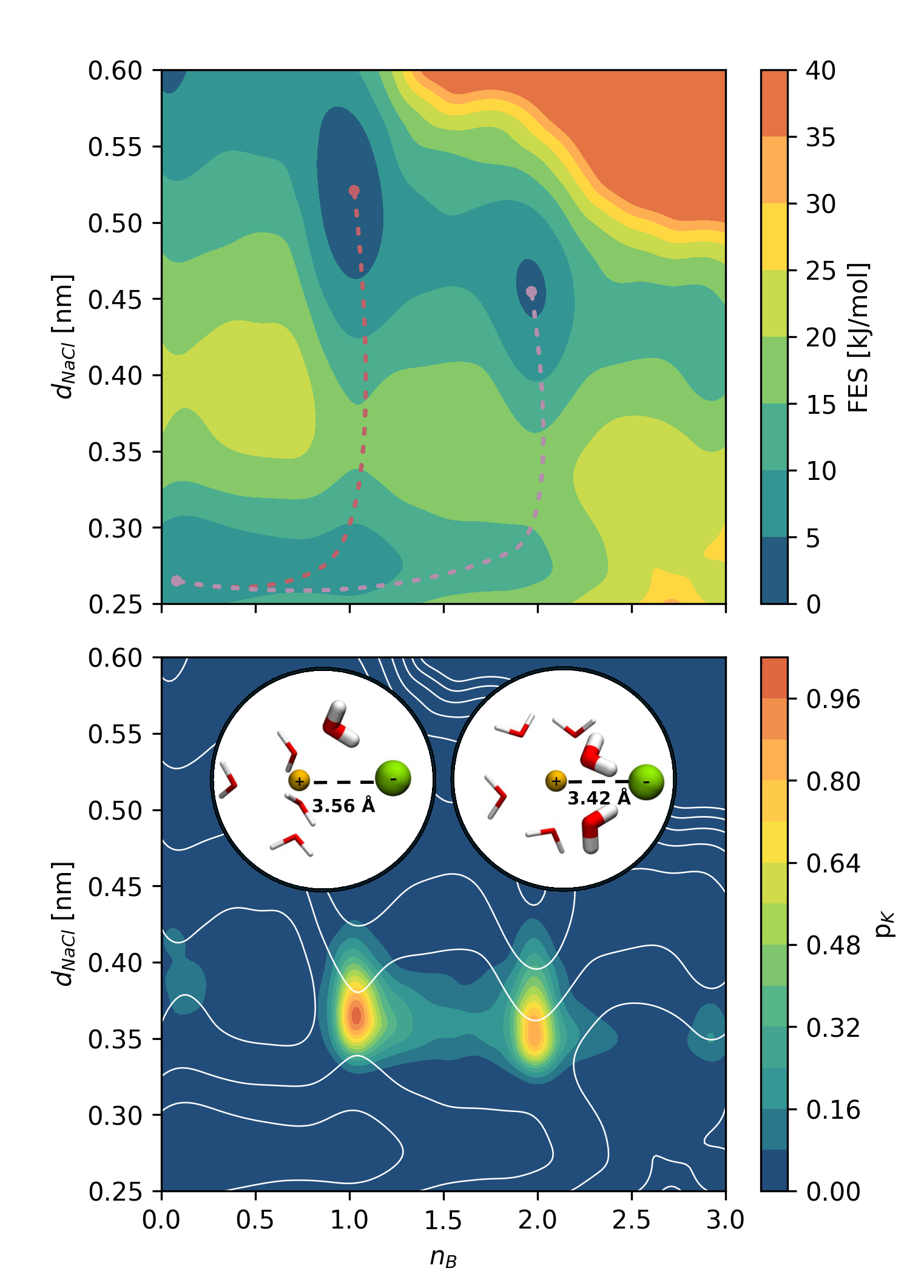}
        \caption{NaCl dissociation. A) Computed free energy surface (FES), indicated by the color map, in the space defined by the number of water molecules bridging the ions $n_b$ and by the inter-ionic distance $d_{NaCl}$.
        The two reactive pathways from the associated states are indicated by dashed lines.
        B) Distribution of the Kolmogorov probability $p_{\mathcal{K}}$, indicated by the color map, in the $n_b$ and $d_{NaCl}$ plane. The isolines of the free energy of A are superimposed in white as a reference.
        The insets show representative configurations of the transition states that characterize the two reaction pathways.}
        \label{fig:NaCl}
    \end{figure}

In Fig.~\ref{fig:NaCl}A, we project the free energy surface onto two physically transparent quantities, the ion-ion distance $d_{NaCl}$ and the number of bridging water molecules that coordinate both ions at the same time $n_B$. 

Usually, the transition state ensemble is defined as the set of  $\boldsymbol{x}$ for which $q (\boldsymbol{x})\approx 0.5$, however, as discussed in Refs.~\citenum{kang2024computing, trizio20241, das2025machine}, it is better to base our analysis on the transition state ensemble as defined by $p_{\mathcal{K}}(\boldsymbol{x})$, which gives the probability that a state $\boldsymbol x$ is actually visited along the reactive trajectory.
In this case, the $p_{\mathcal{K}}(\boldsymbol{x})$ distribution has a multimodal structure in which each mode can be classified by the number of water molecules $n_b$ that are shared by the two ions (see Fig.~\ref{fig:NaCl}B).
Using Eq.~\ref{eq:kolmogorov_distribution} and measuring the integrals over the two main peaks of $p_{\mathcal{K}}(\boldsymbol{x})$, we can estimate that about $50 \%$ of the reactive trajectories pass via the  $n_b\approx 1$ transition state and $40 \%$ via  $n_b\approx 2$.
The remaining 10$\%$ contribution comes from the minority paths through  $n_b\approx 0$ and $n_b\approx 3$.

For this relatively simple system, the above analysis, based on our physical understanding of the system, we could interpret the results rather straightforwardly. 
However, in more complex systems, this may not have been so straightforward.
For this reason, we show that using the information encoded in the GNN, one could have arrived at the same results in an unsupervised way.  
Thus,  we repeated the $p_{\mathcal{K}}(\boldsymbol{x})$ analysis, performing a k-medoid clustering\cite{Schubert2022kmedoids} using as metric the Euclidean distance between the node features of the last GNN layer (see Sec.\ref{sec:interpretability}).
We find again that two modes dominate the $p_{\mathcal{K}}(\boldsymbol{x})$ distribution, which correspond to the two possible screening configurations described above, where the ions are screened by one or two water molecules. 

     \begin{figure*}[t!]
        \centering
        \begin{minipage}{1.0\linewidth}
            \includegraphics[width=\linewidth]{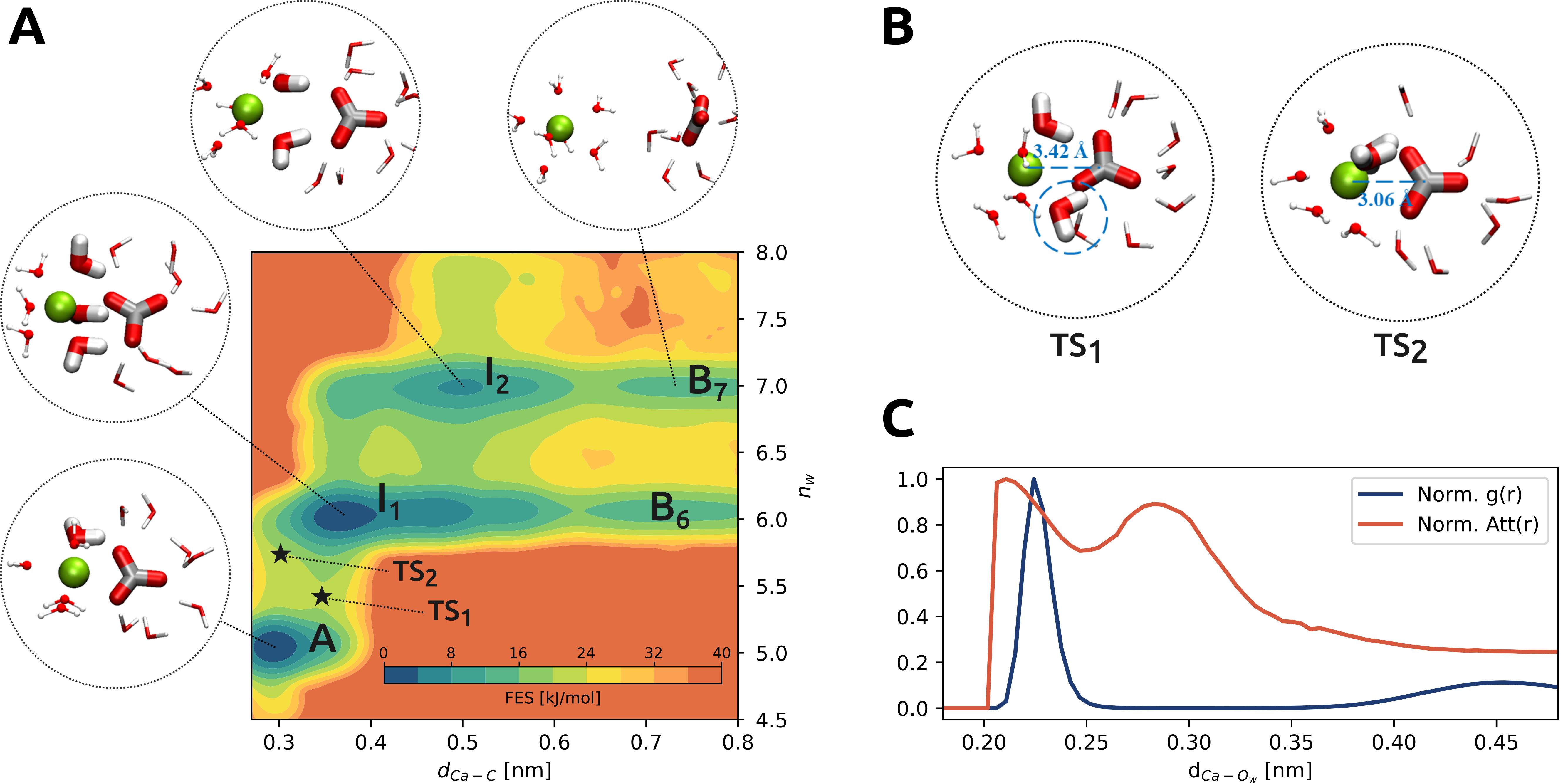} 
        \end{minipage}
        \caption{\ce{CaCO3} dissociation. A) Free energy surface (FES), indicated by the color map, in the plane defined by the inter-ionic distance $d_{Ca-C}$ and the number of solvating water molecules around the \ce{Ca^{2+}} ion $n_w$.
        The metastable states are indicated on the FES, and representative snapshots are provided in the insets.
        The position of the transition states is indicated by a star.
        B) Medoid configurations for the two transition state ensemble clusters TS$_1$ and TS$_2$.
        C) Average attention scores $Att(r)$ of messages from water oxygen nodes to the \ce{Ca^{2+}} node as a function of the interatomic distance $d_{Ca-O_w}$ (red line). 
        The radial distribution function of water oxygens with respect to \ce{Ca^{2+}} is reported as a reference (blue line).
        The two curves are both normalized to be in the range (0,1).}
        \label{fig:CaCO3}
    \end{figure*}

\subsection{\ce{CaCO3}}
 
The final test of our method is the study of the dissociation of \ce{CaCO3} in water, which is another non-trivial example of a chemical process in solution that further showcases the power of the GNN-based approach. 
For this system, we take as state $A$ the contact ion pair in which \ce{Ca^{2+}} and \ce{CO3^{2-}} form a nearly planar structure close to the gas-phase equilibrium geometry, with the \ce{Ca^{2+}} ion symmetrically positioned in front of two carbonate oxygen atoms. 
In such a state, the screening cloud of \ce{Ca^{2+}} is composed by five water molecules and two carboxylic oxygens.
The model potential used at the simulated system size predicts that two major solvation structures ($B_6$ and $B_7$) are possible.
Such structures are almost degenerate: in $B_7$ the cation is surrounded by $7$ water molecules arranged in a pentagonal bi-piramid, while an octahedron of $6$ waters forms the solvation shell of $B_6$.  
The fact that the final state is not unique might at first seem like a major problem for a committor based approach, in which the final state has to be specified beforehand, but in fact using just $B_7$ as final state we were still able to discover $B_6$ and to obtain a free energy surface (see Fig.~\ref{fig:CaCO3}A) in agreement with previous metadynamics-based investigations~\cite{armstrong2023solubility}.  

The resulting FES (see Fig.~\ref{fig:CaCO3}A) shows that, starting from state $A$, the system passes via an intermediate state $I_1$ in which the number of solvating water $n_w=6$. From this state, two reactive paths branch out leading to the two different final states described above.
The intermediate state $I_1$ is characterized by the loss of the planarity characteristic of $A$ and the replacement of one of the coordinating carbonate oxygen atoms by one water molecule.  
From $I_1$, one can go either directly to $B_6$ or to $B_7$ via the intermediate $I_2$ (\( n_w = 7 \)), which corresponds to a near-dissociated configuration in which \ce{Ca^{2+}} is fully coordinated by seven water molecules and the now distant carbonate anion no longer directly contributes to its screening. 
However, in this state, the solvation shells of the two ions still share two water molecules. 

The rate-limiting step of the whole process is the transition from $A$ to $I_1$, which can take place in two ways,
as revealed by the bimodality of the Kolmogorov distribution $P_{\mathcal {K}}(x)$ associated with the transition from A to $I_1$ (see Fig.\ref{SIfig:CaCO3_combine}).
The dominant ensemble of paths passes via $TS_1$ (see Fig. \ref{fig:CaCO3}B), with an interchange mechanism in which the distance between \ce{Ca^{2+}} and \ce{CO3^{2-}} increases, facilitating the arrival of a new water molecule. 
Instead, another less probable path goes via $TS_2$ (see Fig. \ref{fig:CaCO3}B) and can be described as an associative substitution in which both carbonate oxygens remain coordinated, while the calcium-solvating waters temporarily adopt an eight-coordinated geometry.

It is interesting to note that this picture is strengthened by an analysis of the representation learned by the trained GNN. 
In fact, the attention mechanism identifies the water molecules that are most relevant to the \ce{Ca^{2+}} coordination (see Sec.~\ref{sec:methods} for details).  
In Fig.~\ref{fig:CaCO3}C, we plot together, as a function of the  \ce{Ca-O} distance, the pair correlation function 
and the attention weight distribution. 
Unsurprisingly, we see that the primary hydration shell at 2.3 \unit{\angstrom} is identified as highly important, but that a secondary peak at 2.9 \unit{\angstrom} is also relevant. 
This peak is related to the water molecules that take part in the  TS$_1$ ligand-exchange processes, discussed earlier. 
We also observe that the attention distribution starts to be different from zero and rather large as soon as the pair correlation is different from zero, reflecting the fact that the solvation waters at the smaller \ce{Ca-O} distances are more effective at screening the cation and that such an effect is correctly encoded in the attention distribution.

\section{Discussion}

The combination of our recent committor-based approach with the expressivity and generality of the GNN architectures offers a new, powerful tool for the semi-automatic study of rare events.
This approach is descriptor-free, and the graph-based architecture provides new and powerful analysis tools to dissect in detail the reactive process under study.
Such possibilities range from a simpler identification of relevant atoms, thanks to the one-to-one correspondence between reacting atoms and graph nodes, to the use of the model learned hidden representation as the basis for further analysis or the direct study of the information encoded into the attention layers.

Furthermore, we have also shown that the committor iterative optimization procedure\cite{kang2024computing,trizio20241} can be improved and accelerated if one uses data from preliminary imperfect enhanced sampling simulations.
In addition, we believe that our proposed suggestion to reduce the computational cost associated with the use of GNNs for enhanced sampling will help make its routine use more accessible, as it has happened in other fields, such as the construction of interatomic potentials.

\section{Methods} 
\label{sec:methods}
\subsection{Background}

To learn the committor $q(\boldsymbol{x})$, we use the Kolmogorov variational principle \cite{kolmogoroff1931analytischen, weinan2010transition}, which implies minimizing the functional: 
    \begin{equation}
        \mathcal{K}[q(\boldsymbol{x})] =\Big \langle \big|\nabla_{\boldsymbol{u}} q(\boldsymbol{x})\big|^2 \Big\rangle_{U(\boldsymbol{x})} 
        \label{eq:variational_functional}
    \end{equation}
under the boundary conditions $q(\boldsymbol{x}_A)=0$ and $q(\boldsymbol{x}_B)=1$,  where  $\boldsymbol{x}_A$ and $\boldsymbol{x}_B$ denote an initial and final configurations taken from states $A$ and $B$, while $\boldsymbol{u}$ indicates mass scaled coordinates.
The average $\langle \cdot \rangle_{U(\boldsymbol{x})}$ in Eq.\ref{eq:variational_functional} is taken over the Boltzmann ensemble of the studied atomistic system, which we assume to interact via the potential $U(\boldsymbol{x})$. 

However, the evaluation of $\mathcal{K}[q(\boldsymbol{x})]$ is in practice far from trivial.
In fact, the committor has a step-like structure that raises rapidly from $q(\boldsymbol{x}) \approx 0$  when   $\boldsymbol{x}\in A$ to $q(\boldsymbol{x}) \approx 1$ when  $\boldsymbol{x}\in B$,  as it goes through the transition state (TS) region.
This makes the term $\big|\nabla_{\boldsymbol{u}} q(\boldsymbol{x})\big|^2$ sharply peaked on the TS region, which unfortunately is hard to sample, since it is seldom visited when dealing with rare events due to the presence of large energetic barriers.

In order to get around this sampling issue, \cite{kang2024computing, trizio20241}, we thus resort to enhanced sampling. 
In Ref.\citenum{kang2024computing}, to enhance sampling of the otherwise elusive TS region, we introduced the bias potential $V_{\mathcal{K}}(\boldsymbol x)$ 
     \begin{equation}
        V_{\mathcal{K}} (\boldsymbol{x})=-\frac{1}{\beta}\log |\nabla q(\boldsymbol {x})|^2
        \label{eq:bias}
    \end{equation}   
where $\beta$ is the inverse temperature.
Exploiting the above-discussed localization of the committor gradients, such a bias is able to stabilize the TS region and turn it into a minimum that can be sampled as extensively as a standard metastable state.
In addition, to promote also transitions between the states and favor ergodic sampling, in Ref.\citenum{trizio20241}, we complemented this approach with a metadynamics-like bias using the on-the-fly probability enhanced sampling~\cite{invernizzi2020rethinking, trizio20242} (OPES) based on a committor-derived collective variable (CV).
Even if the committor is believed to be formally the best possible CV, its direct use as a CV presents numerical problems that make such an approach ineffective.  
For this reason, we make what basically amounts to the change of variable  $q(\boldsymbol{x}) = \sigma(z(\boldsymbol{x}))$, where $\sigma(z)= 1/\left(1+e^{-pz}\right)$ and use $z(\boldsymbol{x})$ as CV. 
Since $ \sigma(z(\boldsymbol{x}))$ a monotonous function, $z(\boldsymbol{x})$ encodes the same information as $q(\boldsymbol{x})$, but avoids the numerical issues that are a consequence of the sharp behavior of $q(\boldsymbol{x})$ and of the fact that, being $q(\boldsymbol{x})$ a probability, it can assume values that are smaller than the limit of machine precision.

The solution to the Kolmogorov variational problem is found in an iterative process that starts from an initial guess for the committor.  
In Ref.\citenum{kang2024computing}, the initial guess was constructed using data collected in two unbiased simulations performed in the  $A$ and $B$ basins.  
Such an initial guess does not have any information on the TS states, and as such, this initial guess is not optimal.
A better starting guess, which will be used here, is to start the self-consistent procedure, using also data coming also from the TS. Such data can be obtained, for example, from metadynamics simulations, even if driven by suboptimal CVs.

One important feature of the committor is that it provides a powerful analysis tool for understanding the reactive processes. One crucial element in this regard is the identification of those configurations defining the transition state ensemble (TSE).
Following the spirit of our approach, we have proposed to identify the TSE using the Kolmogorov distribution $p_{\mathcal{K}}(\boldsymbol{x})$, defined as
    \begin{equation}
        p_{\mathcal{K}}(\boldsymbol{x}) = 
        \frac{e^{-\beta U_{\mathcal{K}}(\boldsymbol{x}) }}{Z_{\mathcal{K}}} \quad\text{with}\quad U_{\mathcal{K}}(\boldsymbol{x})=U(\boldsymbol{x}) + V_{\mathcal{K}}(\boldsymbol{x})
        \label{eq:kolmogorov_distribution}
    \end{equation}
This distribution indeed measures the contributions that  trajectories passing by a configuration $\boldsymbol{x}$ bring to the transition rate $\nu_R$, since  
        \begin{equation}
            \nu_R \propto \int d \boldsymbol {x} p_{\mathcal{K}}(\boldsymbol{x})  
        \end{equation}
The advantages of using this definition of TSE are discussed in Ref.~\citenum{trizio20241} and in Sec.\ref{sec:results} with examples.

\subsection{Machine learning the committor with GNNs} 

As anticipated in the introduction, we change the model used to parametrize the committor and, rather than using a feed-forward neural network with a set of descriptors as previously done\cite{kang2024computing, trizio20241}, we use a graph neural network (GNN) that directly processes the atomic Cartesian coordinates. 
In addition, we add to the original loss a few regularization terms aimed at stabilizing and simplifying the training.
Apart from these modifications, the procedure is the same as described in our previous work\cite{kang2024computing,trizio20241}.

\paragraphtitle{Loss function}
The loss function employed here is composed of three terms:   
    \begin{equation}
        \mathcal{L} = \log \mathcal{L}_{v} + \alpha_1 \mathcal{L}_{b} + \alpha_2 \mathcal{L}_{r}
        \label{eq:loss_function_new}
    \end{equation}
where the hyper parameters $\alpha_1$ and $\alpha_2$ regulate the relative strength of the different terms.
In the first term, $\mathcal{L}_v$ is the estimate based  on a training  set of   $N_v$ configurations $\boldsymbol{x}_i$  of the functional $\mathcal{K}[q(\boldsymbol{x})]$
    \begin{equation}
        \mathcal{L}_v = \frac{1}{N_v} \sum_{i=1}^{N_v} w_i |\nabla_{\boldsymbol{u}} q(\boldsymbol{x}_i)|^2
        \label{eq:loss_variational}
    \end{equation}  
Since most of the time the data will come from biased simulations, a statistical weight $w_i$ is associated with each $\boldsymbol{x}_i$ so as to give each data its statistically correct contribution.
As in  Ref.\citenum{deng2025role},  we do not insert in the loss function directly  $\mathcal L_v$ but its logarithm to improve training stability since   $\mathcal L_v$ can vary by many orders of magnitude.
This change does not alter the minimum of the functional, since $\mathcal{K}[q(\boldsymbol{x})]$ is positive definite and the logarithm is a monotonously growing function, but avoids numerical problems since $\mathcal{K}[q(\boldsymbol{x})]$  can vary by very many orders of magnitude.

The second term, $L_b$, enforces the correct boundary conditions $q(\boldsymbol{x}_A) = 0$ and $q(\boldsymbol{x}_B) = 1$
    \begin{equation}
        \mathcal{L}_b = \frac{1}{N_A} \sum_{x \in A} (q(\boldsymbol{x}) - 0)^2 + \frac{1}{N_B} \sum_{x \in B} (q(\boldsymbol{x}) - 1)^2,
        \label{eq:loss_boundary}
    \end{equation}
evaluated on a labeled dataset of $N_A$ and $N_B$ configurations belonging to $A$ and $B$  respectively.
Finally, dealing with the greater expressivity of GNN-based models, we realized that it was better to control the range of accessible $z$ values, which in principle could have been $(-\infty,+\infty)$, and avoid overfitting the metastable states.
We thus introduced the regularization term $\mathcal{L}_{r}$ to control the range of the $z$ value accessible and to make optimization more balanced.

\begin{equation}
\mathcal{L}_{r}=\frac{1}{N}\sum_{i=1}^{N}\big[\max(0,\ |z_i|-z_r)\big]^2
\end{equation}
where $z_{r}$ is a user set threshold value and $z_i$  the configurations' predicted $z$ value.

\paragraphtitle{GNN model}
As GNN architecture, we adopt that of SchNet\cite{schutt2018schnet}  proposed by \citeauthor{schutt2018schnet}, which provides a good balance between computational cost and expressivity.
In addition, to better understand the relationship between neighboring nodes, in examples of the \ce{CaCO_3} dissociation and the ligand binding examples, we have applied a node-level attention mechanism to filter out the less relevant messages coming from neighboring nodes.
 
SchNet is a message-passing graph neural network designed for atomistic systems, in which the nodes of the graph represent the atoms in the system.
To each node is also associated a set of features that are initialized based on the corresponding atom type.
Such features are then updated through the network via interaction layers modeled as continuous convolution filters.

In each interaction block, the message passed to the atom $i$ from the atom $j$ is given by
\begin{equation}
    \boldsymbol{m}_{ij} = \boldsymbol{W}(\boldsymbol{h}_j) \, f^{F}_\theta\!\left(\mathsf{RBF}\left(d_{ij}\right)\right)\quad \text{with} \quad d_{ij} = \left|\boldsymbol{x}_i - \boldsymbol{x}_j\right|
\end{equation}

where $\boldsymbol{h}_j$ is the feature array of atom $j$, $\boldsymbol{W}$ is a trainable linear transformation, and $f^{F}_\theta(\cdot)$ is a filter network based on the pairwise interatomic distance expanded by a set of radial basis functions ($\mathsf{RBF}$). The sum of messages from all neighbors is then used to update the representation of atom $i$:  
    \begin{equation}
        \boldsymbol{h}_i^{l+1} = \boldsymbol{h}_i^{l} + f^{M}_\theta\!\left(\frac{1}{N_n} \sum_{j \in \mathcal{N}(i)} \boldsymbol{m}_{ij}\right)
    \end{equation}
where the $j$ index runs over the $N_n$ nodes in the neighborhood $\mathcal{N}(i)$, and $f^{M}_\theta(\cdot)$ is a network that processes the total message received by each node.

Instead, if the attention mechanism is employed, the update function is modified as:
    \begin{equation}
        \boldsymbol{h}_i^{l+1} = \boldsymbol{h}_i^{l} + f^{M}_\theta\!\left(\sum_{j \in \mathcal{N}(i)} \alpha_{ij} \boldsymbol{m}_{ij}\right)
    \end{equation}
where $\alpha_{ij}$ is the trainable attention score~\cite{li2019attention} computed as
    \begin{equation}
        \alpha_{ij} = \frac{exp\left( g_{\theta}(\boldsymbol{m}_{ij}) \right)}{\sum\limits_{k \in \mathcal{N}(i)} exp\left( g_{\theta}(\boldsymbol{m}_{ik}) \right)}
        \label{eqn:atten}
    \end{equation}
in which $g_{\theta}$ is a gate network that returns a scalar value for each message incoming to node $i$.
To summarize, the incorporation of such an attention mechanism enables the network to assign varying importance to neighboring nodes based on their features and spatial relationships, rather than treating all neighbor messages equally.

As anticipated in the previous section, following Ref.~\citenum{trizio20241}, we express the committor as $q(\boldsymbol{x}) = \sigma(z(\boldsymbol{x}))$, with $\sigma$ being the activation function described above.
More in detail, the value of $z$ is computed by first applying a pooling operation (i.e., average) to the node features $\boldsymbol{h}_{i}$ of the last GNN layer and then feeding this to a readout network $f^{R}_\theta$ according to 
    \begin{equation}
        z = f^{R}_{\theta}\left(\frac{1}{N_p} \sum_i^{N_p}\boldsymbol{h}_{i} \right)
    \end{equation}
where $N_p$ is the number of nodes involved in the pooling operation.
This choice proved to be more stable than what was done in our previous work\cite{zhang2024gnncv}, in which the readout function was applied before the pooling operation.
In the most general case, the pooling operation runs on all the nodes in the graph, whereas, in the case of the truncated graph discussed in the next section, the pooling operation is applied to a subset of the most relevant atoms only.

\paragraphtitle{Improving Efficiency} 
An important contribution to the overall efficiency and eventual applicability of our approach comes from the way the model input graph is constructed.
In fact, one of the main limitations of GNN models is their greater computational cost, which scales rapidly with the number of atoms included in the graph.
Often, it is not strictly necessary to include in the graph all the atoms in the system. 
For example, when dealing with chemical reactions in a solvent,  heterogeneous catalysis, or protein ligand binding, only the molecules close to the reactants take an active part in the reaction and should thus be taken into account, whereas the rest of the solvent molecules can be safely ignored. 


To take advantage of these considerations, we use the dual neighbor list schematically depicted in Fig.~\ref{fig:truncated_graph} for the case of a molecule dissociation in water, which is structured as follows.
First, the atoms in the system are divided by the user into two non-overlapping categories. 
The reacting atoms, which are always included in the graph and 
the environment atoms, which are included in the graph only if close to the reacting atoms.
In practice, a first neighbor list $NL_{out}$ is built based on a cutoff $R_t$ with respect to the reacting atoms (panel A), then the graph is constructed by associating nodes to all the atoms included in $NL_{out}$ and connecting them with edges according to a second neighbor list $NL_{graph}$ based on a cutoff $R_c$ (panel B).
Importantly, the cutoffs are chosen  $R_t = R_c + \Delta_b$ to always have a buffer region ($\Delta_b > 0$) of environment atoms not directly interacting with the system, thus stabilizing the calculation between the neighbor lists updates. This is similar to what is done in molecular dynamics when updating neighbor lists.

To avoid atoms belonging to disconnected graphs based solely on $R_c$, as it could easily happen when two ions or a ligand and a guest are far apart in a solvent (panel C), connectivity between reacting atoms can be guaranteed by defining their interaction based on fixed edges between them (panel D). 

The use of this truncated graph approach, already for systems of relatively small sizes, such as the NaCl studied in this paper   (i.e., 216 water molecules),   we find great computational savings both in the training and the simulation stage.

\subsection{GNN Interpretability}
\label{sec:interpretability}

The use of a  geometric GNN to model atomistic systems brings several advantages in terms of the physical insight that can be extracted once the model is optimized.
For example, as atoms are directly mapped to graph nodes, performing a node-level sensitivity analysis can clearly identify which atoms are most relevant for the process in a simple way.
Here, we perform such an analysis in two ways.
In one, we followed the node sensitivity analysis proposed in Ref.\citenum{zhang2024gnncv}, in which the modulus of the derivatives of the model output with respect to the Cartesian coordinates of a node is taken as a measure of its relevance. 
In practice, considering a dataset of $N_g$ graphs, the sensitivity $s_i$ of the $i$-th node is computed as
\begin{equation}
    s_i = \frac{1}{N_g}\sum_{j=1}^{N_g}\left|\frac{\partial z\left(\mathcal{G}^j\right)}{\partial{\boldsymbol{x}}_{i}^{j}}\right| 
    \label{eq:node_level_sensitivity}
\end{equation}
Where, $\mathcal{G}^j$ is the $j$-th entry of the dataset, and $\boldsymbol{x}_{i}^{j}$ is the position of the $i$-th node in $\mathcal{G}^j$.

Alternatively, when an attention layer is employed, one can use the attention scores $\alpha_{ij}$ as a measure of node relevance of the $i,j$  interaction (see Eq.\ref{eqn:atten}). 

Another important advantage of GNN models is that, starting from simple atomic coordinates, they inherently learn a representation of the system that encodes structural information in the hidden node features of the model. 
In fact, a distance metric, for instance, the Euclidean distance, can be defined in the latent space identified by such features that can be used to perform unsupervised analysis, for example, based on clustering algorithms or other dimensionality reduction algorithms.

%% file: manuscript/supporting.tex
\section{Details about the SchNet-GNN based on truncated graph.}


As we have introduced in the main text, to reduce the computational cost of the GNN used in our committor model, we designed the so-called truncated graph method, which predict the committor value only using atoms near the reaction center of the investigated process.
Briefly speaking, during the construction of the graph, we first build an ``atom cluster'' composed of ``system'' atoms and ``environment'' atoms within a specified radius from the reacting atoms, then generate input graph data using only atoms within such a truncated atom cluster.
In the training stage, this procedure is performed for each entry in the dataset, whereas during MD propagation, we dynamically update the indices of the environment atoms included in the cluster.
As a result, the number of atoms within the graph can fluctuate during simulation.
To avoid sudden changes in the committor value caused by environment atoms entering or leaving the graph, we build a buffer region, as often done with neighbor lists.
In practice, we truncate the system using a radius slightly larger than the graph cutoff radius ($R_c$), extended by a buffer size ($\Delta_b$).
Therefore, when an entering or leaving event occurs, the corresponding environment atom is not linked to the reacting atoms by any direct edge.

However, even with such a procedure, entering or leaving atoms will nevertheless contribute to the committor value, since there are still edges between such atoms and other environment atoms inside the graph.
Thus, we carried out the following two modifications to the SchNet-GNN.
First, we added a cosine cutoff function $f^c$ to the internode message calculation process, and the new message function reads as follows:
\begin{equation}
\boldsymbol{m}_{ij} = \boldsymbol{W}(\boldsymbol{h}_j)f^{F}_\theta\!\left(\mathsf{RBF}\left(d_{ij}\right)\right)f^c\!\left(d_{ij}\right) \quad \text{with} \quad
 f^c(d) =
\begin{cases} 
\left[ 1 + cos\left(\pi d/R_c\right) \right]/2, & d < R_c \\
0. & d \geq R_c
\end{cases}
\label{eq:cos_cutoff}
\end{equation}
Then, more importantly, we used only the features of the system nodes in the pooling function to predict the committor value:
\begin{equation}
z = f^{R}_{\theta}\left(\frac{1}{N_s} \sum_{i\in \mathcal{S}}\boldsymbol{h}_{i}\right),
\end{equation}
where $\mathcal{S}$ stands for the set of system nodes and $N_s$ is the number of system nodes.
Through the message-passing mechanism, when the GNN model has more than one layer, the system node features still carry information from the environment atoms and therefore reflect the overall structure of the truncated atom cluster.
However, due to the action of the cutoff function $f^c$, environment atoms that are far from reaction atoms will contribute less to the model output.
Especially, messages from atoms inside the buffer region will be weakened twice before they are merged into the system node features, and thus influence even less the model output.
That is, by employing both the decay function and the selective pooling operation, we implemented the truncated graph-based GNN models in a smooth manner.

\begin{figure}[h!]\centering
\includegraphics[width=0.5\linewidth]{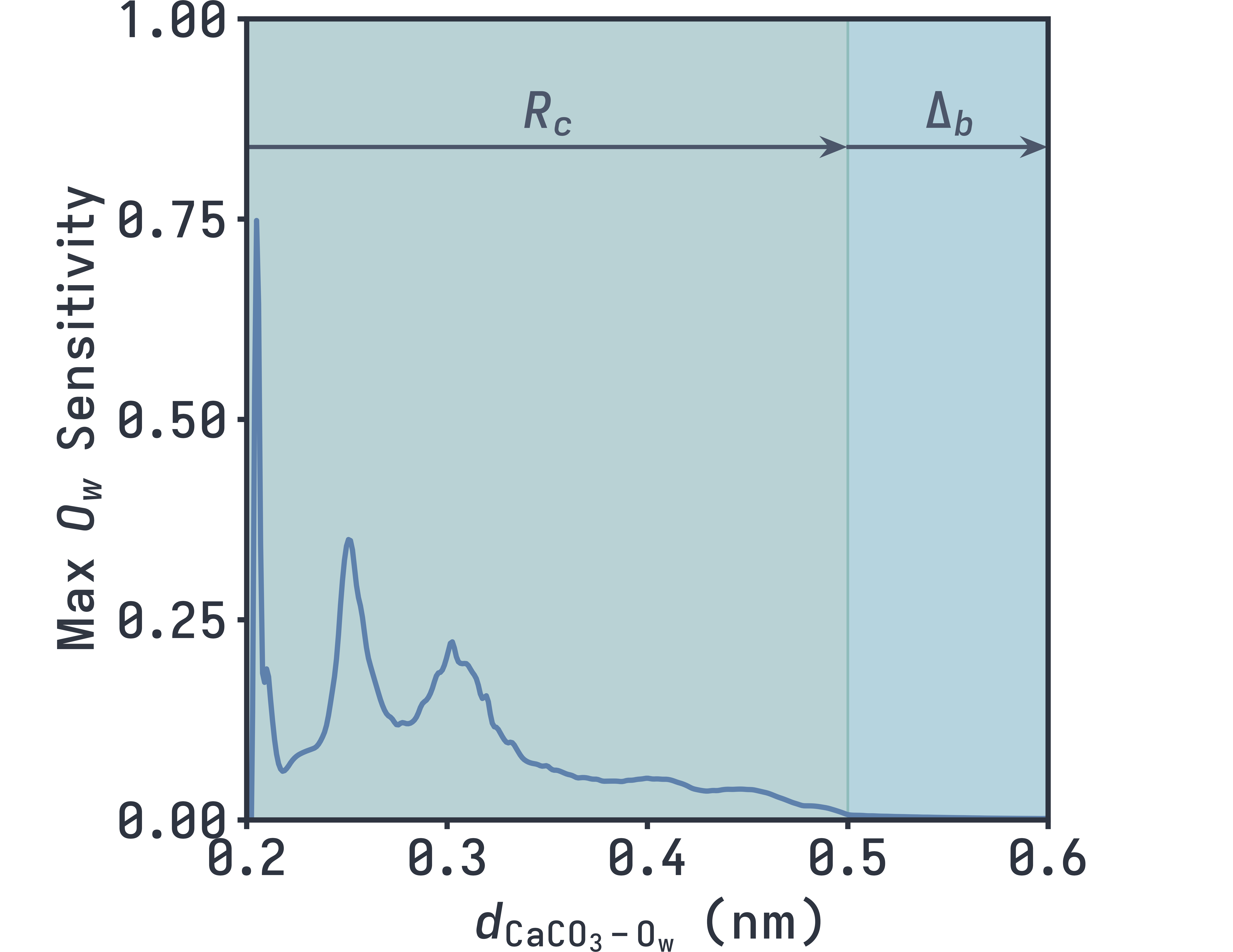}
\caption{Maximum sensitivity of water oxygen atoms in the \ce{CaCO3} system as a function of their shortest distances from the ion pair.}
\label{fig:CaCO3_d_s}
\end{figure}

To demonstrate the action of the above procedure, we plot the maximum sensitivity of water oxygen atoms in the \ce{CaCO3} system as a function of their distances from the solute in Fig.~\ref{fig:CaCO3_d_s}.
As can be seen, as the solute-solution distance approaches the graph cutoff distance, positions of the corresponding solution atom contribute less and less to the predicted committor value.
For solution atoms inside the buffer region, their sensitivity indicates that they are in fact negligible to the committor model.
As a result, the implemented GNN-based committor can be truly smooth when a solution atom enters or leaves the truncated graph, and thus is suitable for biased enhanced sampling simulations.

%




\clearpage
\section{Alanine Dipeptide - Additional information}
    \subsection{Computational details}
        \paragraphtitle{Simulations details}
        We utilized the AMBER99SB~\cite{amber2013} force field to describe the alanine dipeptide molecule (Ace-Ala-Nme) in a vacuum.
        The lengths of all chemical bonds involving hydrogen atoms were constrained at their equilibrium values using the LINCS~\cite{hess1997lincs} algorithm.
        The electrostatic and Lennard-Jones (LJ) interactions were described in a non-cutoff manner.
        All MD propagations were carried out under the \textit{NVT} ensemble using the GROMACS v2021.5~\cite{abraham2015gromacs} package with the Langevin integrator~\cite{kieninger2022sd}, in combination with the PLUMED~\cite{tribello2014plumed, plumed} plugin.
        The integration time step and the friction coefficient were set to 2 \unit{\fs} and 20 \unit{\ps^{-1}}, correspondingly.

    \paragraphtitle{Committor model training details}
        To model the committor function \(q_\theta(\boldsymbol{x})\) at each iteration, we employed all 10 heavy atoms as inputs to the graph neural network. Since the system does not include solvent, no distinction between core and environment atoms was necessary here.
        The graph cutoff radius $R_c$ is selected as 1.0 \unit{\nm}.
        The GNN architecture consisted of three layers, with 16 Gaussian basis functions, an output feature dimensionality of 20, and a two-layer feed-forward readout network. For message aggregation, we used the minimum-value operation.
        
        For the optimization, we used the ADAM optimizer with an initial learning rate of $4*10^{-4}$ modulated by an exponential decay with multiplicative factor $\gamma=0.99993$. 
        The training was performed for 5000 epochs in the first iteration and for $\sim$40000 epochs for the others. 
        The $\alpha$ hyperparameter in the loss function was set to 10. 
        The number of iterations, the corresponding dataset size, and the $\lambda$ and the OPES \texttt{BARRIER} used in the biased simulations are summarized in Table~\ref{sup_tab:alanine_iterations} alongside the lowest value obtained for the functional $K_m$, which provides a quality and convergence measure, the simulation time $t_s$ and the output sampling time $t_o$. To test the robustness of the GNN architecture, in this example, we used only unbiased data for the initial guess and did not employ the fast-converging OPES+$V_{\mathcal{K}}$ strategy; instead, OPES and $V_{\mathcal{K}}$ biased trajectories were run separately.

            \begin {table}[h!]
                \caption {Summary of the iterative procedure for Alanine.} \label{sup_tab:alanine_iterations}
                \begin{center}
                \begin{tabular}{ |c|c|c|c|c|c|c| } 
                 \hline
                 Iteration & Dataset size & $K_m$ [au] & OPES \texttt{BARRIER} [kJ/mol] & $\lambda$ & $t_s$ [ns] & $t_o$ [ps] \\ 
                 \hline
                    0   & 20000 & 10180  & - & -   & 2*10 & 1 \\
                    1   & 50000 & 132  & 35 & 0.8  & 2*20 & 1 \\
                    2   & 80000 & 10.3 & 35 & 0.8  & 2*20 & 1 \\ 
                    3   & 110000 & 2.19 & 35 & 0.8  & 2*20 & 1 \\     4  &  140000 & 2.09 & 35 & 0.8  & 2*20 & 1 \\
                    5   & 170000 &  1.75  & 35 & 0.8 & 2*20 & 1 \\
                 \hline
                \end{tabular}
                \end{center}
            \end {table}

    \subsection{Additional figures}
\begin{figure*}[h!]
    \centering
    \begin{minipage}{0.5\linewidth}
        \includegraphics[width=\linewidth]{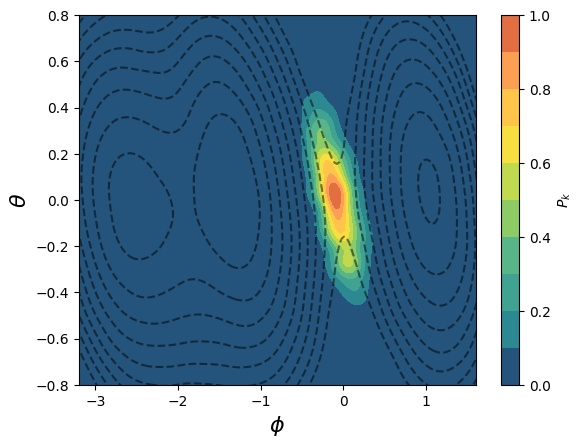} 
    \end{minipage}
    \caption{\textbf{Additional information of the committor of alanine dipeptide}: $p_{\mathcal{K}}$ distribution projected on $\phi$ and $\theta$.}
    \label{SIfig:alad_pk}
\end{figure*}

\clearpage
\section{Calixarene - Additional information}
    \subsection{Computational details}
        \paragraphtitle{Simulations details}
            We utilized the Amber General Force Field (GAFF)~\cite{wang2004development} with the RESP~\cite{bayly1993well} charges and the TIP3P~\cite{jorgensen1983tip3p} water model to describe the Calixarene system.
            In our setup, a pair of OAMe host molecule and \ce{G2} guest molecule was solvated at a density of about 1 \unit{g/cm^3} with 2100 water molecules in a cubic box of 4.027 $\times$ 4.027 $\times$ 4.027 \unit{\nm^3}.
            Nine extra sodium ions are included to counterbalance excess charges. 
            The lengths of all chemical bonds involving hydrogen atoms were constrained at their equilibrium values using the LINCS~\cite{hess1997lincs} algorithm.
            Electrostatic interactions were described using the Particle Mesh Ewald (PME) \cite{essmann1995} method with a real-space cutoff at 1.0 \unit{\nm}.
            The Lennard-Jones (LJ) interactions were calculated with a distance cutoff of 1.0 \unit{\nm}.
            All MD propagations were carried out under the \textit{NVT} ensemble using the GROMACS v2021.5~\cite{abraham2015gromacs} 
            with the Langevin integrator~\cite{kieninger2022sd} in combination with the PLUMED~\cite{tribello2014plumed, plumed} plugin.
            The integration time step and the friction coefficient were set at 2 \unit{\fs} and 1 \unit{\ps^{-1}}, correspondingly.
            At every simulation step, the coordinates are aligned so that the vertical axis of the box coincides with the binding axis $h$, and the simulation box is centered on the virtual atom \ce{V1}.
    
        \paragraphtitle{The funnel restraint}
        \label{sup_sec:funnel_restraint}
            In our simulations, we used a funnel restraint~\cite{limongelli2013funnel} equivalent to the one previously employed by Refs.\citenum{rizzi2021,bhakat2017resolving,perez2019local} on the same system.
            Here, we summarize the details of such a restraint, while more details can be found in the original works.
            The funnel limits the space available to the ligand in the unbound state U by confining it to a cylindrical volume above the binding site. 
            As the ligand approaches the binding site, the funnel restraint becomes wider so that its presence does not affect the binding process itself. 
            Having aligned with PLUMED the system to a reference configuration where the binding axis is found along the vertical axis, we define $h$ as the projection on the binding axis of the center of the carbon atoms of each ligand and $r$ as its radial component.
            When $h>10 \si{\angstrom}$, the funnel surface is a cylinder with radius $R_{cyl} = 2 \si{\angstrom}$ with its axis along the vertical direction. 
            When $h<10 \si{\angstrom}$, the funnel opens into an umbrella-like shape with a 45 degree angle whose surface is defined by $r=12-h$.
            The force that, for a displacement x, pushes the ligand away from the funnel’s surface is harmonic $-k_Fx$ with $kF=20 $ kJ/mol $\si{\angstrom}^{-2}$. 
            A further harmonic restraint is applied on $h$  to prevent the ligand from getting too far from the host, reaching the upper boundary of the simulation box. 
            The corresponding force is $-k_U(h-18)$ for $h > 18\si{\angstrom}$ and $k_U=40 $ kJ/mol $\si{\angstrom}^{-2}$.

            During training, we set boundaries further to state U so that the labeled configurations used to impose the boundary conditions will not influence the committor training in the subsequent iterative simulations.
            We apply the funnel restraint described above and two additional harmonic restraints $-k_U(h-20)$ for $h > 20 \si{\angstrom}$ and $-k_U(h-18)$ for $h < 18 \si{\angstrom}$, with $k_U = 20$ kJ/mol.
        
            Because of the funnel presence, the free energy difference between the bound and the true unbound state that we extract from enhanced sampling simulations needs a correction that can be calculated as:
                \begin{equation}
                    \Delta G = -\frac{1}{\beta} \log \left( C_0 \pi R_{\mathsf{cyl}}^2 \int_B dh \exp\left[-\beta \left(W(h) - W_U\right)\right] \right)
                \end{equation}
            where $\beta$ = 1/($k_B$T), $C_0$ = 1/1660 $\si{\angstrom}^{-3}$ is the standard concentration, $h$ is the coordinate along the funnel’s axis, $W(h)$ is the free energy along the funnel axis and $W_U$ its reference value in state U. 
            More precisely, we define $W_U$ as the average free energy value in the interval 1.6  $\si{\angstrom}$ < $h$ < 1.8  $\si{\angstrom}$.
            The integral is computed over the state B region that we define as 0.3  $\si{\angstrom}$ < $h$ < 0.8  $\si{\angstrom}$.

\begin{figure*}[h!]
        \centering
        \begin{minipage}{0.3\linewidth}
            \includegraphics[width=\linewidth]{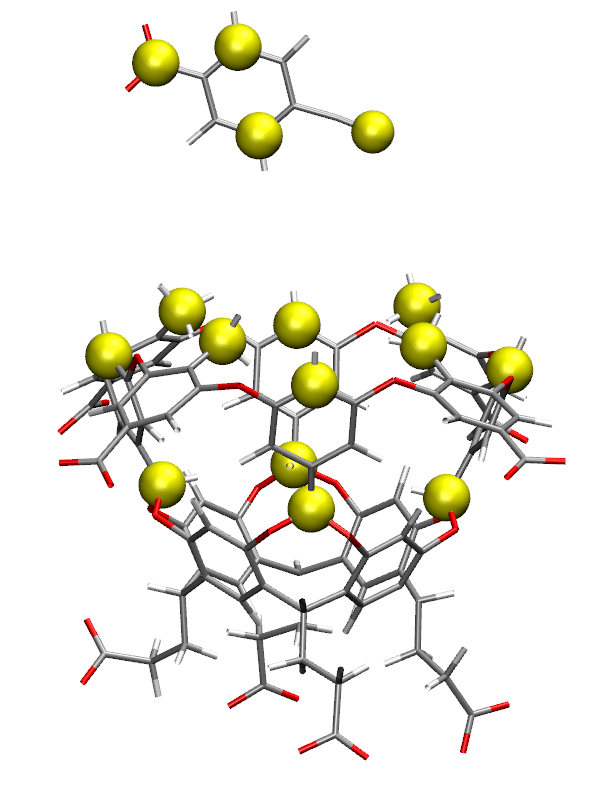} 
        \end{minipage}
        \caption{\textbf{Calixarene atoms set in the GNN architecture}: the yellow atoms represent the selected reacting atoms}
        \label{SIfig:Calixarene_coreatoms}
    \end{figure*}

        \paragraphtitle{Committor model training details}
            To model the committor function $q_\theta(\boldsymbol{x})$ at each iteration, we used 4 atoms on the guest molecule and 12 atoms on the host skeleton as the reacting atoms of the system(see the labeled yellow atoms in Fig \ref{SIfig:Calixarene_coreatoms}), and the neighboring oxygen atoms of water as environmental atoms. The cutoff radius were set to \(R_c = 5 \,\unit{\angstrom}\) and \(R_b = 6 \,\unit{\angstrom}\). 
            The GNN architecture consisted of two layers, with 12 Gaussian basis functions, an output feature dimensionality of 16, and a two-layer feed-forward readout network. For message aggregation, we used the attention mechanism described in the Method part.

        For the optimization, we used the ADAM optimizer with an initial learning rate of $8*10^{-4}$ modulated by an exponential decay with multiplicative factor $\gamma=0.9999$. 
        The training was performed for 1000 epochs in the first iteration and for $\sim$3000 epochs for the others. 

            \begin {table}[h!]
                \caption{Summary of the iterative procedure for Calixarene.} \label{sup_tab:calixarene_iterations}
                \begin{center}
                \begin{tabular}{ |c|c|c|c|c|c|c| } 
                 \hline
                 Iteration & Dataset size & $K_m$ [au] & OPES \texttt{BARRIER} [kJ/mol] & $\lambda$ & $t_s$ [ns] & $t_o$ [ps] \\ 
                 \hline
                    0   & 100000 &2250  & 50 & - & 100 & 1 \\
                    1   & 120000 & 254.9 & 50 & 0.6 & 100 & 1 \\  
                    2   & 120000 &23.5  & 50 & 0.6 & 100 & 1 \\
                    3   & 120000 &159 & 50 & 0.6 &   100 & 1 \\ 
                    4   & 120000 & 5.99 & 50 & 0.6 & 100 & 1 \\  
                    5   & 120000 & 8.11 & 50 & 0.6 & 100 & 1 \\  
                 \hline
                \end{tabular}
                \end{center}
            \end {table}

    \clearpage

    \subsection{Additional results}

        \paragraphtitle{Free energy surface}

            \begin{figure*}[h!]
        \centering
        \begin{minipage}{0.6\linewidth}
            \includegraphics[width=\linewidth]{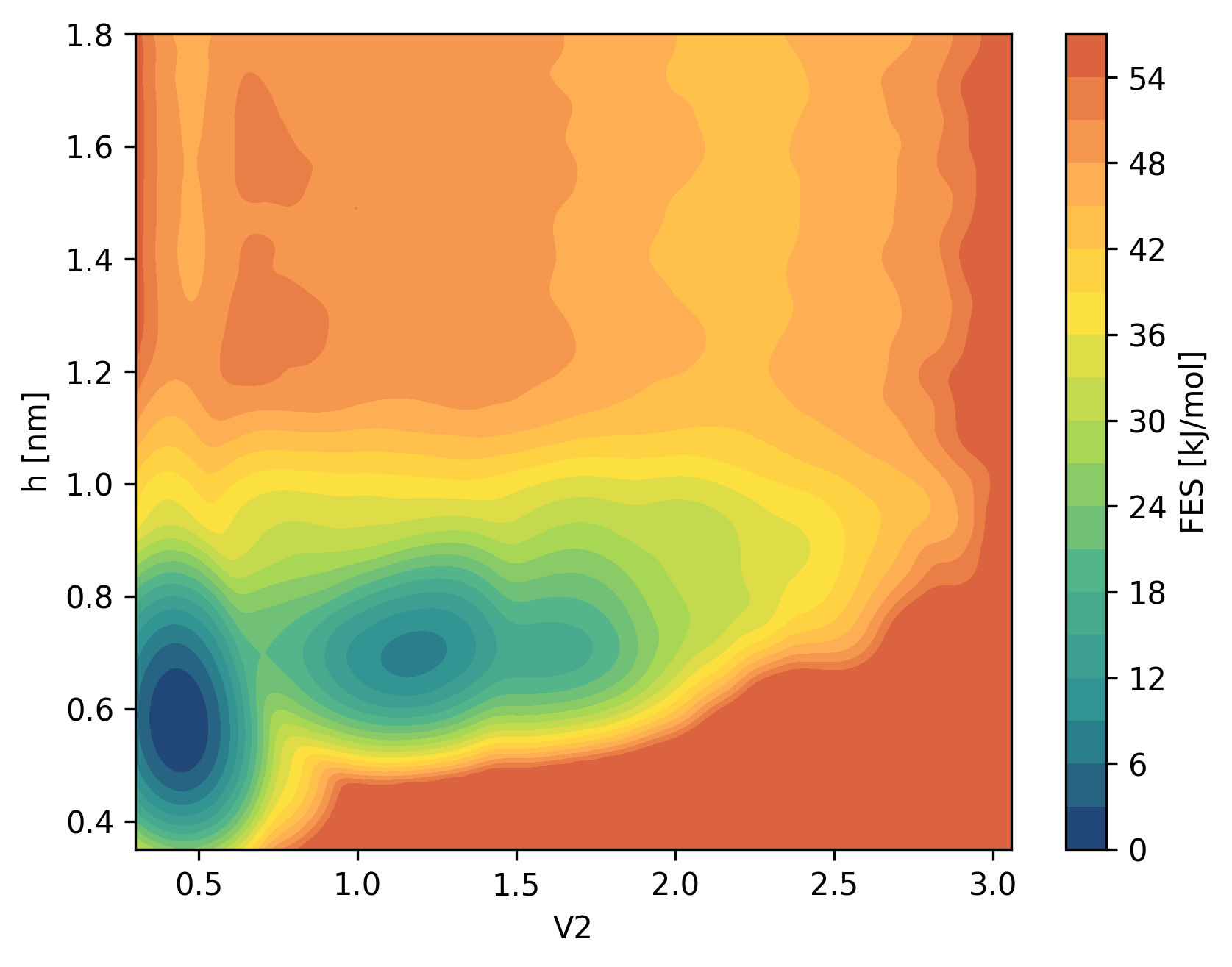} 
        \end{minipage}
        \caption{\textbf{2D free energy surface calixarene} Two-dimensional free energy surface for calixarene projected in the plane defined by the water coordination number of a virtual point inside the binding cavity (\ce{V2}) and the projection of the ligand molecule on the binding axis (h). }
        \label{SIfig:Calixarene_2DFES}
    \end{figure*}

    \paragraphtitle{ $\Delta F$ Convergency}
    
    \begin{figure*}[h!]
        \centering
        \begin{minipage}{0.6\linewidth}
            \includegraphics[width=\linewidth]{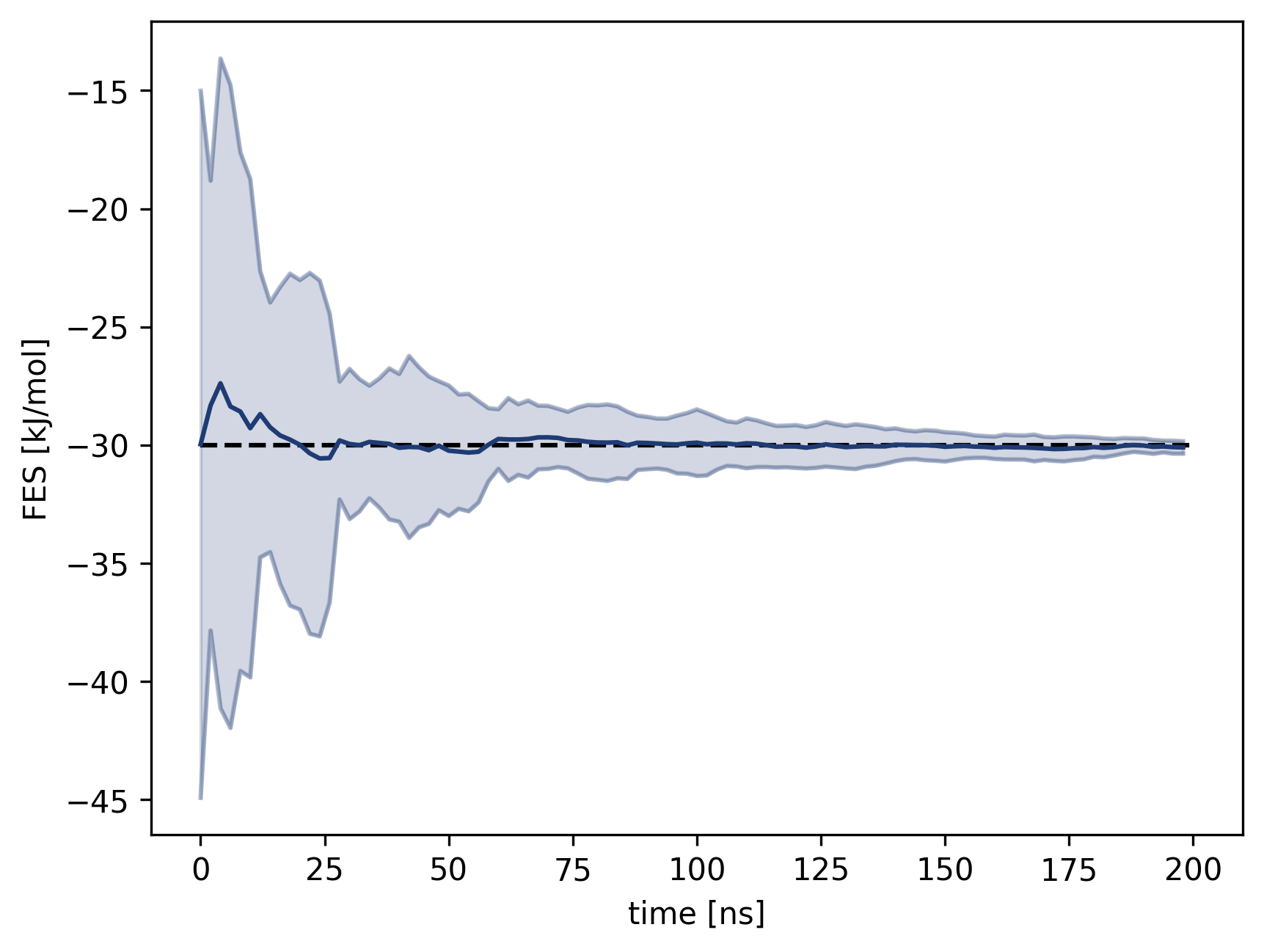} 
        \end{minipage}
        \caption{\textbf{Convergence with simulation time of the binding energy of the \ce{G2} ligand to the OAMe octa-acid guest.} The average estimates from 3 independent simulations are reported as a blue solid line, whereas the uncertainty, computed as the standard deviation over the three replicas, is depicted as a shaded blue region. The reference values are provided as a black dashed line, and the reference estimate is obtained using the enhanced sampling setup of Ref.~\citenum{rizzi2021}}
        \label{SIfig:Calixarene_deltaF}
    \end{figure*}

    \paragraphtitle{The snapshots of two pathways}
        
    \begin{figure*}[h!]
        \centering
        \begin{minipage}{1.0\linewidth}
            \includegraphics[width=\linewidth]{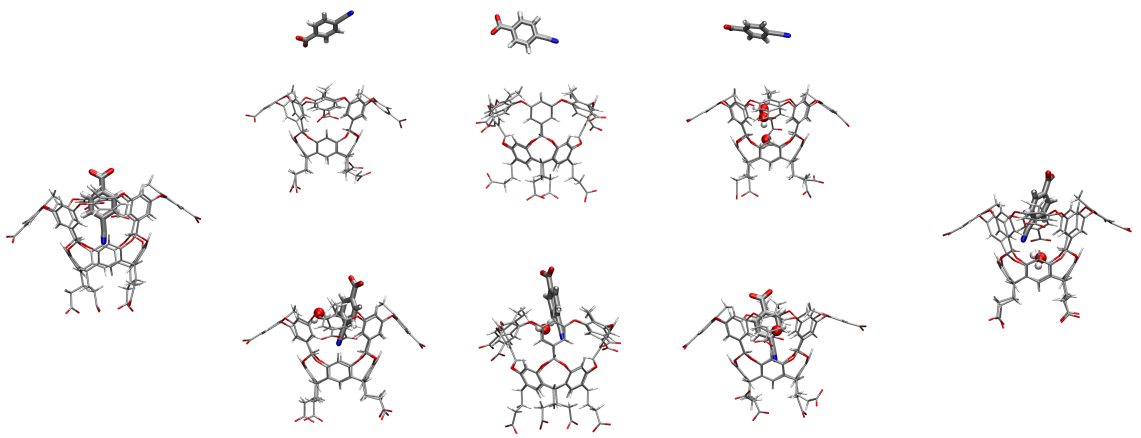} 
        \end{minipage}
        \caption{\textbf{Calixarene snapshots}
        Snapshots of representative configurations along the reaction pathway of $B_d$ to $B_w$, grouped according to the $z$ value. Water molecules inside the pocket are highlighted, while external waters are omitted for clarity. The upper pathway corresponds to a highly unlikely theoretical transition via the unbinding state, whereas the lower pathway represents the dominant route discussed in the main text. }
        \label{SIfig:Calixarene_snapshots_twopath}
    \end{figure*}

\clearpage
\section{N\MakeLowercase{a}C\MakeLowercase{l} - Additional information}
    \subsection{Computational details}
        \paragraphtitle{Simulations details}
        We utilized the force field parameters proposed by \citeauthor{joung2008ff}~\cite{joung2008ff} along with TIP3P~\cite{jorgensen1983tip3p} waters to describe the \ce{NaCl} solution system.
        In our setup, a pair of \ce{NaCl} ions was solvated at a density of about 1 \unit{g/cm^3} with 216 water molecules in a cubic box of 1.86 $\times$ 1.86 $\times$ 1.86 \unit{\nm^3}.
        The lengths of all \ce{H-O} bonds were constrained at their equilibrium values using the RATTLE~\cite{andersen1983} algorithm.
        Electrostatic interactions were described using the Particle Mesh Ewald (PME) \cite{essmann1995} method with a real-space cutoff at 0.9 \unit{\nm}.
        The Lennard-Jones (LJ) interactions were calculated with a distance cutoff of 0.9 \unit{\nm},
        All MD propagations were carried out under the \textit{NVT} ensemble using the OpenMM~\cite{openmm} package with the Geodesic BAOAB Langevin integrator~\cite{leimkuhler2016} provided by OpenMMTools,~\cite{openmmtools} in combination with the PLUMED~\cite{tribello2014plumed, plumed} plugin.
        The integration time step and the friction coefficient were set at 2 \unit{\fs} and 1 \unit{\ps^{-1}}, correspondingly.
        The number of geodesic drift steps was set to 4.

        \paragraphtitle{Committor model training details}\label{sup_sec:Nacl training}
            To model the committor function $q_\theta(\boldsymbol{x})$ at each iteration, we used two ions \ce{Na+} and \ce{Cl-} as the reacting atoms of the system and the neighboring oxygen atoms of water are considered as environment atoms.
            The graph cutoff radius $R_c$ is 0.6 \unit{\nm} and the buffer size $\Delta_b$ is 0.1 \unit{\nm}.
            The GNN architecture consisted of three layers, with 14 Gaussian basis functions, an output feature dimensionality of 24, and a two-layer feed-forward readout network. For message aggregation, we used the minimum-value operation.
            For the optimization, we used the ADAM optimizer with an initial learning rate of $4*10^{-4}$ modulated by an exponential decay with multiplicative factor $\gamma=0.9999$. 
            The training was performed for 1000 epochs in the first iteration and for $\sim$2000 epochs for the others.

            \begin {table}[h!]
                \caption {Summary of the iterative procedure for \ce{NaCl}} \label{sup_tab:nacl_iterations}
                \begin{center}
                \begin{tabular}{ |c|c|c|c|c|c|c| } 
                 \hline
                 Iteration & Dataset size & $K_m$ [au] & OPES \texttt{BARRIER} [kJ/mol] & $\lambda$ & $t_s$ [ns] & $t_o$ [ps] \\ 
                 \hline
                    0   & 38000 & 3.25 & - & -   & 5 & 1 \\
                    1   & 68000 &  2.72  & 15 & 0.2 & 16 & 0.5 \\
                 \hline
                \end{tabular}
                \end{center}
            \end {table}

            \begin {table}[h!]
                \caption {Summary of the iterative procedure for \ce{NaCl} starting from only unbiased data} \label{sup_tab:nacl_iterations}
                \begin{center}
                \begin{tabular}{ |c|c|c|c|c|c|c| } 
                 \hline
                 Iteration & Dataset size & $K_m$ [au] & OPES \texttt{BARRIER} [kJ/mol] & $\lambda$ & $t_s$ [ns] & $t_o$ [ps] \\ 
                 \hline
                    0   & 10000 & 23.3 & - & -   & 5 & 1 \\
                    1   & 40000 &  3.13  & 15 & 0.2 & 16 & 0.5 \\
                    2   & 70000 &  2.89  & 15 & 0.2 & 16 & 0.5 \\
                 \hline
                \end{tabular}
                \end{center}
            \end {table}

\clearpage
\section{C\MakeLowercase{a}\ce{CO3} - Additional information}
    \subsection{Computational details}
        \paragraphtitle{Simulations details}
        We utilized the force field parameters for both ions and water proposed by \citeauthor{armstrong2023ff}~\cite{armstrong2023ff} to describe the \ce{CaCO3} solution system.
        In our setup, a pair of \ce{CaCO3} ions was solvated at a density of about 1 \unit{g/cm^3} with 818 water molecules in a cubic box of 2.9011 $\times$ 2.9011 $\times$ 2.9011 \unit{\nm^3}.
        Electrostatic interactions were described using the Particle Mesh Ewald (PME) \cite{essmann1995} method with a real-space cutoff at 0.9 \unit{\nm}.
        The Lennard-Jones (LJ) interactions were calculated with a distance cutoff of 0.9 \unit{\nm}.
        Specifically, the long-range dispersion correction to the LJ interactions was not applied, as required by the force field~\cite{armstrong2023ff}.
        All MD propagations were carried out under the \textit{NVT} ensemble using the OpenMM~\cite{openmm} package with the BAOAB Langevin integrator~\cite{leimkuhler2013} provided by OpenMMTools,~\cite{openmmtools} in combination with the PLUMED~\cite{tribello2014plumed, plumed} plugin.
        The integration time step and the friction coefficient were set at 1 \unit{\fs} and 1 \unit{\ps^{-1}}, correspondingly.

        \paragraphtitle{Committor model training details}

            To model the committor function \(q_\theta(\boldsymbol{x})\) at each iteration, we defined the ions \ce{{Ca}^{2+}} and \ce{CO3^{2-}} as the reacting atoms of the system, while the neighboring oxygen atoms of water were treated as environment atoms. The graph cutoff radius $R_c$ is 0.6 \unit{\nm} and the buffer size $\Delta_b$ is 0.1 \unit{\nm}. In addition, to account for the large separation between ions upon full dissociation, we imposed fixed edges between the \ce{Ca^{2+}} cation and all atoms belonging to the \ce{CO3^{2-}} anion.
            The GNN architecture consisted of two layers, with 12 Bessel basis functions, an output feature dimensionality of 16, and a two-layer feed-forward readout network. For message aggregation, we used the attention mechanism described in the Method part.

            For the optimization, we used the ADAM optimizer with an initial learning rate of $10^{-3}$ modulated by an exponential decay with multiplicative factor $\gamma=0.9999$. 
            The training was performed for 1000 epochs in the first iteration and for $\sim$4000 epochs for the others. 

            \begin{table}[h!]
                \caption {Summary of the iterative procedure for \ce{CaCO_3}.} \label{sup_tab:caco3_iterations}
                \begin{center}
                \begin{tabular}{ |c|c|c|c|c|c|c| } 
                 \hline
                 Iteration & Dataset size & $K_m$ [au] & OPES \texttt{BARRIER} [kJ/mol] & $\lambda$ & $t_s$ [ns] & $t_o$ [ps] \\ 
                 \hline
                    0   & 45000 &4.99  & 25 & - & 30 & 1 \\
                    1   & 34000 &3.92 & 25 & 0.4 & 15 & 1 \\ 
                    2   & 68000 & 2.90 & 25 & 0.4 & 30 & 1 \\  
                 \hline
                \end{tabular}
                \end{center}
            \end {table}
    \subsection{Additional results}        
\begin{figure*}[h!]
    \centering
    \begin{minipage}{1.0\linewidth}
        \includegraphics[width=\linewidth]{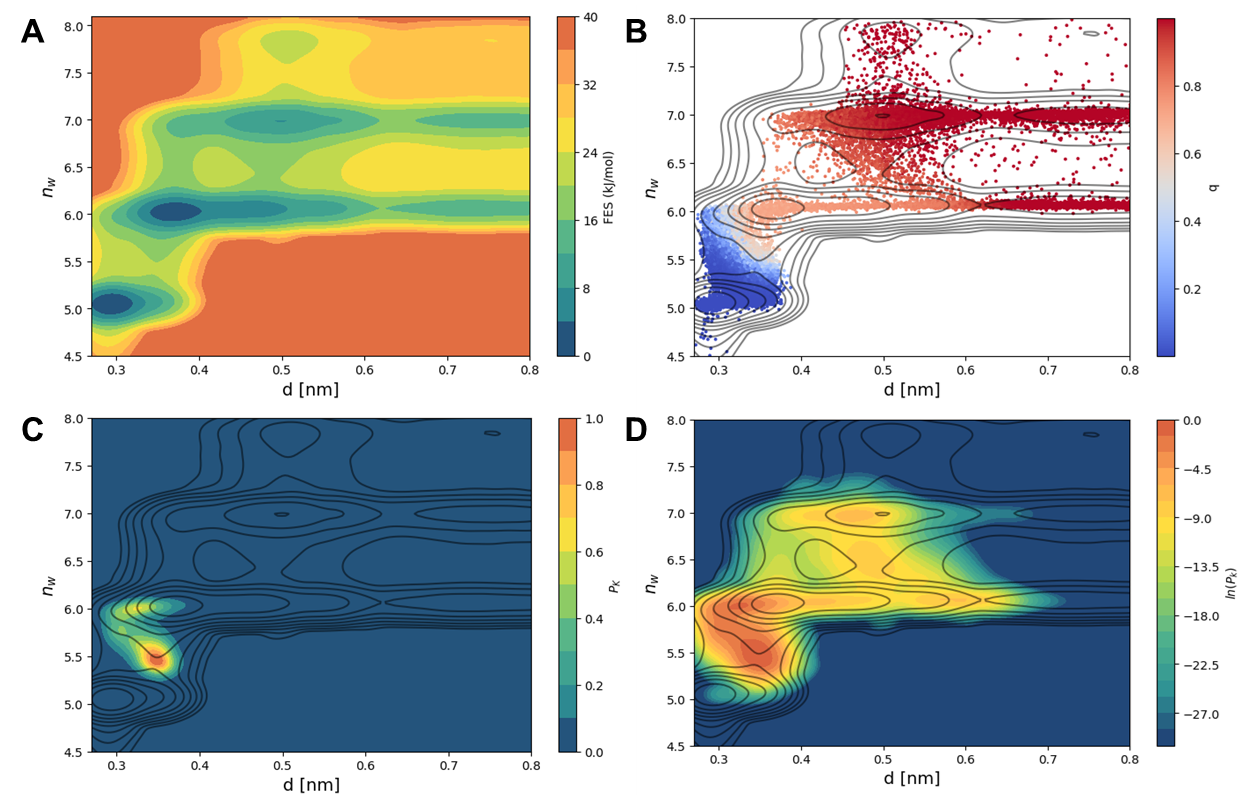} 
    \end{minipage}
    \caption{\textbf{Additional information of the committor of \ce{CaCO3} dissociation}:A) 2D FES projected on d and $n_w$ B) Simulated dataset colored by committor value. C) $p_{\mathcal{K}}$ distribution projected on d and $n_w$. D) $log(p_{\mathcal{K}})$ distribution projected on d and $n_w$. to show more details of the $p_{\mathcal{K}}$ distribution in the second barrier}
    \label{SIfig:CaCO3_combine}
\end{figure*}

\begin{figure*}[h!]
    \centering
    \begin{minipage}{0.6\linewidth}
        \includegraphics[width=\linewidth]{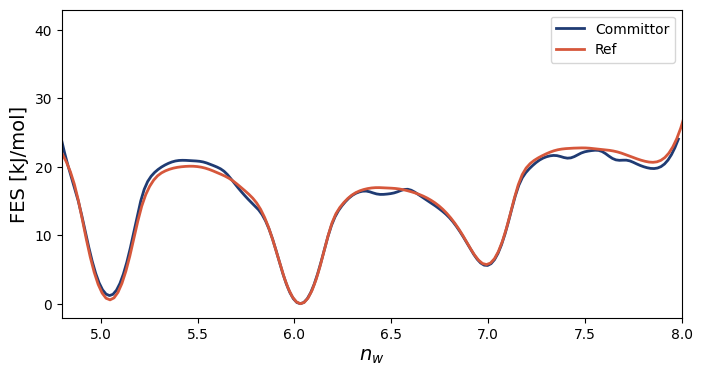} 
    \end{minipage}
    \caption{\textbf{1D FES of \ce{CaCO3} dissociation process projected on $n_w$}: The committor estimate (blue) is obtained from our OPES+$V_{\mathcal{K}}$ simulation biased on committor, and the reference estimate(red) is obtained from a long $OPES$ simulation biased on $d$ and $n_w$}
    \label{fig:CaCO3_1D}
\end{figure*}